\documentclass[11pt]{amsart}

\usepackage{amssymb,amsthm,amsmath}
\usepackage[numbers,sort&compress]{natbib}
\usepackage{color}
\usepackage{graphicx}
\usepackage{tikz}
\usepackage[colorlinks,
            linkcolor=blue,
            anchorcolor=blue,
            citecolor=blue
            ]{hyperref}

\newcommand{\SL}{\mathrm{SL}(2,\mathbb{R})}

\let\newpf\proof \let\proof\relax 
\newenvironment{pf}{\newpf[\proofname]}{\qed\endtrivlist}

\newcommand{\ba}{\overline{A}}

\def\be{\begin{equation}}
\def\ee{\end{equation}}

\def\ba{{\begin{align}}}
\def\ea{{\end{align}}}

\def\bm{\begin{matrix}}
\def\em{\end{matrix}}

\def\u{{\mathbb U}}

\def\SL{{\mathrm{SL}}}

\def\0{{\mathbf 0}}

\newtheorem{Theorem}{Theorem}[section]
\newtheorem{Lemma}{Lemma}[section]
\newtheorem{Proposition}{Proposition}[section]

\newtheorem{Remark}{Remark}[section]

\newtheorem{Definition}{Definition}[section]

\numberwithin{equation}{section}

\theoremstyle{definition}

\def\tr{{\text{tr}}}

\renewcommand{\mod}{\operatorname{mod}}

\newcommand{\id}{\operatorname{id}}

\newcommand{\N}{{\mathbb N}}
\newcommand{\Q}{{\mathbb Q}}
\newcommand{\R}{{\mathbb R}}
\newcommand{\T}{{\mathbb T}}

\newcommand{\Z}{{\mathbb Z}}

\def\B0{{\bold{0}}}

\catcode`\@=12

\def\Empty{}
\newcommand\oplabel[1]{
  \def\OpArg{#1} \ifx \OpArg\Empty {} \else
    \label{#1}
  \fi}

\newcommand{\comm}[1]{}
\newcommand{\comment}[1]{}

\begin{document}

\title{Second Phase transition line}

\author{Artur Avila}
\address{CNRS UMR 7586, Institut de Math\'ematiques de Jussieu -
Paris Rive Gauche, B\^atiment Sophie Germain, Case 7012, 75205 Paris
Cedex 13, France \& IMPA, Estrada Dona Castorina 110, 22460-320, Rio
de Janeiro, Brazil}
 \email{artur@math.jussieu.fr}

\author {Svetlana Jitomirskaya}
\address{
Department of Mathematics, University of California, Irvine CA, 92717
} \email{szhitomi@uci.edu}

\author{Qi Zhou}
\address{CNRS UMR 7586, Institut de Math\'ematiques de Jussieu -
Paris Rive Gauche, B\^atiment Sophie Germain, Case 7012, 75205 Paris
Cedex 13, France}
\curraddr{Department of Mathematics, Nanjing University, Nanjing 210093, China
} \email{qizhou628@gmail.com,qizhou@nju.edu.cn}

\date{\today}

\begin{abstract}
We study the  phase transion line of the almost
Mathieu operator, that separates arithmetic regions corresponding to
singular continuous and a.e. pure point regimes, and prove that both purely singular continuous and  a.e. pure point spectrum occur
for dense sets of frequencies. 
\end{abstract}

\setcounter{tocdepth}{1}

\maketitle

\section{Main results}

In systems with phase transitions the interface between the
two phases often exhibits the critical phenomena and is the most
difficult set of parameters to study. At the same time, the insights on the critical
case often shed light on the creation, dissipation, and the mechanism behind
both  phases. In this paper we study the critical regime for the
hyperbolic almost Mathieu operator.

 It has been  known since the work of Aubry-Andre \cite{AA80} that  the almost Mathieu family:
\begin{equation}\label{schro}
(H_{\lambda,\alpha,\theta} u)_n= u_{n+1}+u_{n-1} +2\lambda \cos 2
\pi (n\alpha + \theta) u_n,
\end{equation}
where $\theta\in \mathbb{R}$ is  the phase, $\alpha\in \R\backslash
\Q$ is the frequency and $\lambda\in \R$ is  the coupling constant,
undergoes a phase transition at $\lambda=1,$ where the Lyapunov
exponent changes from zero everywhere on the spectrum \cite{BJ} to
positive everywhere on the spectrum \cite{H}. Aubry-Andre conjectured
\cite{AA80}, that at $\lambda=1$
 the spectrum changes from absolutely continuous for $\lambda<1$
to pure point for $\lambda>1.$ This has since been proved, for all
$\alpha,\theta$ for $\lambda<1$ \cite{L93,Aab,AD,AJ08}  and for Diophantine $\alpha,\theta$
(so a.e.) for $\lambda>1$ \cite{J99}. The ``a.e.'' cannot be removed as
in the hyperbolic regime, while there is no absolutely continuous
spectrum, the distinction between singular continuous and pure point
depends in an interesting way on the
arithmetics of  $\alpha,\theta.$ The relevant issue is an interplay
between the rate of exponential growth of the transfer-matrix cocycle
and the depth of the small denominators: the exponential rate of
approximation of $\alpha$ by the rationals. A conjecture dating back
to 1994 \cite{J95}  (see also
\cite{J07})  was that for a.e. $\theta$ there is another
transition, from singular continuous to pure point spectrum, precisely
where the two rates compensate each other. Namely, for $\alpha\in \R\backslash
\Q$ 
with continued fraction approximants  $\frac{p_n}{q_n}$, let
$$\beta(\alpha):=\limsup_{n\rightarrow \infty}\frac{\ln
q_{n+1}}{q_n}.$$ $\beta(\alpha)$ measures how exponentially
Liouvillean $\alpha$ is. Then at $\lambda=e^{\beta}$ there is, for
a.e. $\theta$, a
transition from singular continuous to pure point
spectrum.\footnote{The exclusion of a measure zero set is not needed
  for the singular continuous part, but is necessary for the pure
  point part \cite{JS}.} This was
proved recently in \cite{AYZ} by an improvement of the Gordon-type
method for the singular continuous region and reducibility as a
corollary of subcriticality and duality for the pure point one. Moreover, the arithmetic version for the pure point region
(specifying the a.e. $\theta$ as phases $\theta$ that are
$\alpha$-Diophantine) was established in \cite{JL}, through a
constructive proof of localization, that demonstrated also the continued
fraction expansion driven universal hierarchical structure of
corresponding eigenfunctions. Finally, for $\lambda=1$ there is
singular continuous spectrum except  for an explicit full measure set of
$\theta$ (namely, for $\theta$'s which are not rational with respect to $\alpha$) \cite{AK06,AJM,G,L94}. We can summarize those known results in
\begin{Theorem}\label{AYZ}
Let $\alpha\in \R\backslash \Q$, then we have the following:
\begin{enumerate}
\item If $ |\lambda|<1,$ then $H_{\lambda,\alpha,\theta}$ has
purely absolutely continuous spectrum for all $\theta$.
\item If  $ |\lambda|=1,$ then $H_{\lambda,\alpha,\theta}$ has
purely singular continuous spectrum for (explicit) a.e. $\theta$.
\item  If $1\leq |\lambda|<e^\beta,$ then $H_{\lambda,\alpha,\theta}$ has
purely singular continuous spectrum for all $\theta$.
\item  If $|\lambda|>e^\beta,$ then $H_{\lambda,\alpha,\theta}$ has
purely point spectrum with exponentially decaying eigenfunctions
for (explicit) a.e. $\theta$.
\end{enumerate}

\end{Theorem}
Therefore, as far as a.e. $\theta$ is concerned only the phase
transition case
$|\lambda|=e^\beta, 0<\beta<\infty,$ is missing. This is what we want to address in
this paper.  We  will show

\begin{Theorem}\label{main}
Let $0<\beta<\infty.$ Then

\begin{enumerate}
\item  There exists a dense set of $\alpha$  with $\beta(\alpha)=\beta$ such
  that  $H_{\lambda,\alpha,\theta}$
  with  $\lambda=e^{\beta(\alpha)}$, has
purely singular continuous spectrum for all $\theta$.
\item  There exists a dense set of $\alpha$  with $\beta(\alpha)=\beta$ such
  that  $H_{\lambda,\alpha,\theta}$ with  $\lambda=e^{\beta(\alpha)}$, has
pure point spectrum 
for a.e. $\theta$.
\end{enumerate}

\end{Theorem}

\begin{Remark}
\begin{enumerate}
\item We note that $\lambda=e^{\beta(\alpha)}$ implies that there are no
exponentially decaying eigenfunctions \cite{JL}. Thus part $(2)$ provides an example of
nonexponentially decaying eigenfunctions in the regime of positive Lyapunov
exponents. 
\item The ``a.e.'' in part $(2)$ cannot be improved to ``all'' \cite{JS}.
\item 
  As usual, $\ln \lambda$ can be viewed as a shortcut for Lyapunov exponent $L(E)$ on the
  spectrum of the almost Mathieu operator, so it is natural to
  conjecture that for general analytic potentials there will be a
  transition at  $L(E)=\beta(\alpha)$.  Indeed,  singular continuous spectrum does
 hold throughout the $L(E)<\beta(\alpha)$ regime even for Lipschitz
  potentials. \footnote{This is essentially contained in \cite{AYZ}. 
 Additionally, it follows from a more recent theorem of \cite{jy} where
  singularities are also allowed.}
  However, the localization result for general analytic case does
  not currently exist even for the Diophantine case (nor under any
  other arithmetic condition). Moreover, by Avila's global theory \cite{Aglobal} Lyapunov
  exponent is a stratified analytic function with finitely many
  strata. Thus the set $A_{cr}:=\{E: L(E)=\beta(\alpha)\}$
  will only be uncountable if $L$ is constant on one of the
  strata. However, even for the potentials where the set $A_{cr}$ is
  a small subset of the spectrum, the study of what happens at those
  energies is still interesting as they represent the border between
  two different behaviors.
\end{enumerate}
\end{Remark}

\section{Preliminaries}

Let $\T =\R/\Z.$ For a bounded
analytic (possibly matrix valued) function $F$ defined on $ \{ \theta |  | \Im \theta |< h \}$, let
$
\|F\|_h=  \sup_{ | \Im \theta |< h } \| F(\theta)\| .$ Let
$C^\omega_{h}(\T,*)$ be the
set of all these $*$-valued functions ($*$ will usually denote $\R$,
$SL(2,\R)$). Also we denote $C^\omega(\T,*)=\cup_{h>0}C^\omega_{h}(\T,*)$.

\subsection{Continued Fraction Expansion}\label{sec:2.1}
Let $\alpha \in (0,1)$ be irrational. Define $ a_0=0,
\alpha_{0}=\alpha,$ and inductively for $k\geq 1$,
$$a_k=[\alpha_{k-1}^{-1}],\qquad
\alpha_k=\alpha_{k-1}^{-1}-a_k=G(\alpha_{k-1}):=\{{1\over
  \alpha_{k-1}}\}.$$ 
%
Let $p_0=0,  p_1=1,  q_0=1,  q_1=a_1,$ and we define inductively
$$p_k=a_kp_{k-1}+p_{k-2},\qquad q_k=a_kq_{k-1}+q_{k-2}.$$
Then $(q_n)$  is the sequence of  denominators of the best rational
approximations of $\alpha$ and we have \begin{equation}\label{appro1} \forall 1
\leq k < q_n,\quad \|k\alpha\|_{\T} \geq \|q_{n-1}\alpha\|_{\T},
\end{equation}
and
\begin{equation}\label{appro2}
\frac{1}{2q_{n+1}} \leq \|q_n \alpha \|_{\T} \leq {1 \over q_{n+1}}.
\end{equation}

As a direct consequence of  $(\ref{appro1})$ and $(\ref{appro2})$, we have the  following:
\begin{Lemma}\label{bdc}
Suppose that there exists $p\in\N$, such that $a_i=1$ if $i\geq p.$ Then we  have 
\begin{eqnarray}
\|k \alpha \|_{\T} \geq \frac{1}{4|k|},      \qquad \forall  |k| \geq q_{p-1} .
\end{eqnarray}
\end{Lemma}

For any $\alpha, \alpha^{'}\in \R\backslash \Q$
, let  $n$ be  the
first index for which the continued fraction expansions of $\alpha$
and $\alpha^{'}$ differ. Define $d_{H}(\alpha, \alpha^{'})=
\frac{1}{n+1}.$ Then 
$(\R\backslash \Q,d_H)$
is a complete metric
space. Also, by $(\ref{appro2})$, if $d_{H}(\alpha, \alpha^{'})=
\frac{1}{n+1}$, we have that $|\alpha-\alpha^{'}|<
\frac{1}{q_{n-1}(\alpha)^2}.$  

Finally, we introduce the set of $\alpha$-Diophantine phases. For  $\tau>1,$
$\gamma>0,$ set
$$DC_\alpha(\gamma,\tau)=\{ \phi\in \R| \|2\phi-m\alpha\|_{\R/\Z}\geq
\frac{\gamma}{(|m|+1)^\tau}, m\in\Z\},$$
Clearly, $\cup_{\gamma>0}DC_\alpha(\gamma,\tau)$ is a full measure set.

\subsection{Cocycle, reducibility,  rotation number}

Let $\alpha \in \R \backslash \Q$, $A \in C^\omega(\T$, $\SL(2,\R))$.
We define the quasi-periodic $SL(2,\R)$-cocycle $(\alpha,A)$ as an
action on  $\T\times \R^2$ by $(x,v)\rightarrow (x+\alpha,A(x)v).$
Recall that  two cocycles $(\alpha, A^{i}),i=1,2,$ are called $C^k$ $(k=\infty,\omega)$
conjugated if there exists  $B\in C^k(\T,$ $PSL(2, \mathbb{R}))$ such
that
$A^{1}(\theta)=B(\theta+\alpha)A^{2}
(\theta)B(\theta)^{-1}.$ If $(\alpha, A)$ is $C^k$ conjugated to
a constant cocycle, then it is called $C^k$ \textit{reducible}.  A cocycle
$(\alpha, A)$ is said to be \textit{almost reducible} if
the closure of its analytical conjugacy class contains a constant.

 Assume now that $A:\T \to \SL(2,\R)$ is homotopic to the identity.  Then
there exist $\psi:\T \times \T \to \R$ and $u:\T\times \T \to
\R^+$ such that
\begin{equation}
A(x) \cdot \left (\bm \cos 2 \pi y \\ \sin 2 \pi y \em \right )=u(x,y)
\left (\bm \cos 2 \pi (y+\psi(x,y)) \\ \sin 2 \pi (y+\psi(x,y)) \em \right
).
\end{equation}
The function $\psi$ is called a {\it lift} of $A$.  Let $\mu$ be any
probability measure on $\T \times \T$ which is invariant under the continuous
map $T:(x,y) \mapsto (x+\alpha,y+\psi(x,y))$, projecting over Lebesgue
measure on the first coordinate.  Then the number
\begin{equation}
\rho(\alpha,A)=\int \psi d\mu \mod \Z
\end{equation}
does not depend on the choices of $\psi$ and $\mu$, and is called the
{\it fibered rotation number} of
$(\alpha,A)$, see \cite {H} and \cite {JM82}.

Rotation number plays a fundamental role in the reducibility theory:
\begin{Theorem}\cite{AFK,HoY}\label{hy1} Let
$(\alpha, A) \in \R\backslash \Q \times C^\omega_{h}(\T,SL(2,\R))$
with  $h>h'>0,$ $R\in SL(2,\R)$. Then for every $\tau>1,$
$\gamma>0,$ if $\rho(\alpha, A)\in DC_\alpha(\gamma,\tau),$
then there exists $\varepsilon=\varepsilon(\tau,\gamma,  h-h' ),$ such that if  $\|
 A(\theta)-R\|_h<\varepsilon(\tau,\gamma,  h-h' )$, then there exist $B \in C^\omega_{h'}(\T,SL(2,\R))$,
$\varphi \in C^\omega_{h'}(\T,\R)$,
 such that
$$B(\theta+\alpha) A(\theta) B(\theta)^{-1}= R_{\varphi(\theta)}.$$
Moreover, we have the following estimates
\begin{enumerate}
\item $\|B-\id\|_{h'}\leq \|
 A(\theta)-R\|_h^{\frac{1}{2}}$, 
\item $\|\varphi(\theta)-\hat\varphi(0)\|_{h'}\leq 2\|
 A(\theta)-R\|_h.$
\end{enumerate}
\end{Theorem}

\section{Singular continuous spectrum}
Before  giving the proof of Theorem \ref{main}(1), we introduce
some notations. Denote the spectrum of $H_{\lambda,\alpha,\theta}$ by
$\Sigma(\lambda,\alpha).$ It doesn't depend on $\theta$, since $\theta \rightarrow \theta+\alpha$ is minimal. Let $$A(E, \theta)=S_{E}^{\lambda}(\theta)=\left( \begin{array}{ccc}
 E-2\lambda\cos2\pi (\theta) &  -1\cr
  1 & 0\end{array} \right),$$ $A_0=I,$ and for $k\geq1$ we set
\begin{eqnarray*}A_k(E,\theta)&=&A(E,\theta+(k-1)\alpha)\cdots
  A(E,\theta+\alpha)A(E,\theta),\\
A_{-k}(E,\theta)&=&A_k(E,\theta-k\alpha)^{-1}. 
\end{eqnarray*} 
In case both $\alpha$ and $\alpha'$ are involved in the argument, we
will use

$$A^{'}_k(E,\theta)=A(E,\theta+(k-1)\alpha^{'}  )\cdots
A(E,\theta+\alpha^{'})A(E,\theta)$$ and $A'_{-k}(E,\theta)=A'_k(E,\theta-k\alpha')^{-1}. $
A formal solution $u_E^{\theta}(n)$ of
\begin{equation}\label{ev}H_{\lambda,\alpha,\theta}u_E^{\theta}(n)=Eu_E^{\theta}(n) \end{equation}
is said
to be normalized if $|u_E^{\theta}(0)|^2+|u_E^{\theta}(1)|^2=1$. We
will call such solutions normalized eigenfunctions of
$H_{\lambda,\alpha,\theta}.$ 

We now introduce the following concept:

\begin{Definition}
For any $C>0$, $N \in \N$, we say  $(\lambda,\alpha) $ is $(C,N)$ bad,
if for any $\theta\in \T$, $E\in \Sigma(\lambda,\alpha)$, and for any
normalized solution $u_E^{\theta}(n)$ of (\ref{ev}),
we have  $ \sum_{|k|\leq N} |u_E^{\theta}(k)|^2$ $\geq C^2$. 
\end{Definition}

We then have 
\begin{Lemma}\label{rem}
Let $\lambda>1$. If for any $C>0$ there exists $N\in\N$ such that $(\lambda,\alpha)$ is  $(C,N)$ bad,  then $H_{\lambda,\alpha,\theta}$ has purely singular continuous spectrum for any $\theta\in\R$.
\end{Lemma}
\begin{pf}
The assumption of Lemma \ref{rem} implies absence of $\ell^2$
solutions. Since Lyapunov exponents are positive for all $E$ for
$\lambda>1$ \cite{H}, there is no absolutely continuous spectrum either \cite{Ko84}.
\end{pf}
Of course the notion of $(C,N)$ badness is very general and can be defined for an
arbitrary potential with obvious modifications. Similarly, Lemma
\ref{rem} is also a very general statement, requiring only positivity
of the Lyapunov exponent. The converse is not true in general: 
purely singular continuous spectrum (or absence of point spectrum) for all phases doesn't necessarily
imply $(C,N)$ badness for any $C.$ However in our case, we do have the following:
\begin{Proposition}\label{gordonayz}
If $1\leq |\lambda|<e^\beta,$ then for  any $C>0$, there exists $N\in\N$ such that $(\lambda,\alpha)$ is  $(C,N)$ bad.
\end{Proposition}

\begin{pf}
This is a Gordon-type statement that is essentially contained in the proof of
Theorem \ref{AYZ} \cite{AYZ}. We just give a short argument here for
completeness.  If $1\leq |\lambda|<e^\beta,$ then for any
$\epsilon>0$,  by uniform upper semicontinuity and telescoping (see e.g. Proposition 3.1 of  \cite{AYZ}), one has the following:  there exists $K=K(\lambda, \alpha,\epsilon)$ which doesn't depend on $\theta,E$,  such that if $n\geq K$, we have 
\begin{eqnarray}
\label{appro-3}\sup_{\theta\in\T}\| A_{q_n}(E,\theta+q_n\alpha)-
A_{q_n}(E,\theta)\|& \leq& 
e^{-(\beta-\ln\lambda- \epsilon)q_n}.\\
\label{appro-4}\sup_{\theta\in\T}\| A_{-q_n}(E, \theta+q_n\alpha)-
A_{-q_n}(E,\theta)\|& \leq& e^{-(\beta-\ln\lambda- \epsilon)q_n}.
\end{eqnarray}
The following lemma  essentially completes the proof: 
\begin{Lemma} Suppose that $(\ref{appro-3})$ and $(\ref{appro-4})$
  hold,  and  let  $u_E^{\theta}(k)$ be a normalized solution of (\ref{ev}). Then we have 
\begin{equation}\label{refinegordon} \max\{ \|A_{q_{n}}(E,\theta)\overline{u}_E^{\theta}\|, \|A_{-q_{n}}(E,\theta)\overline{u}_E^{\theta}\|, \|A_{2q_{n}}(E,\theta)\overline{u}_E^{\theta}\|
 \} \geq \frac{1}{4},\end{equation}
 where  $\overline{u}_E^{\theta}= \begin{pmatrix} u_E^{\theta}(0)\\u_E^{\theta}(1)
\end{pmatrix} $.
\end{Lemma} 
\begin{pf}
For any $M\in SL(2,\R)$, one has
\begin{equation}\label{hami-cal}
  M+M^{-1}=\tr M\cdot Id.
  \end{equation} Taking $M=A_{q_n}(E,\theta)$, we distinguish two cases. \\
\textbf{Case 1:} If $|\tr A_{q_n}(E,\theta)| >1$, $(\ref{hami-cal})$ gives 
\begin{equation}\label{refinegordon-1} \max\{ \|A_{q_{n}}(E,\theta)\overline{u}_E^{\theta}\|, \|A_{q_{n}}(E,\theta)^{-1}\overline{u}_E^{\theta}\| \} \geq \frac{1}{2}.\end{equation}
Then $(\ref{refinegordon})$ follows from $(\ref{refinegordon-1})$ and $(\ref{appro-4})$.\\
\textbf{Case 2:} If $|\tr A_{q_n}(E,\theta)| <1$, then by $(\ref{hami-cal})$, one has 
\begin{eqnarray*}
&&  \|A_{2q_{n}}(E,\theta)\overline{u}_E^{\theta}\|\\
&=& \| A_{2q_n}(E,\theta)\overline{u}_E^{\theta}-
A_{q_n}(E,\theta) ^2\overline{u}_E^{\theta}-  \overline{u}_E^{\theta}+\tr A_{q_n}(E,\theta) \cdot    A_{q_{n}}(E,\theta)\overline{u}_E^{\theta} \| \\
&\geq & 1- (1+ \| A_{q_n}(E,\theta+q_n\alpha)-
A_{q_n}(E,\theta)\|)\|A_{q_{n}}(E,\theta)\overline{u}_E^{\theta}\|
\end{eqnarray*}
Then $(\ref{refinegordon})$ follows from  $(\ref{appro-3})$.
\end{pf}

This directly implies that for any $C>0$,  $(\lambda,\alpha)$ is  $(C,N)$ bad for sufficiently large $N$. 
 \end{pf}

The notion of $(C,N)$-badness is robust with respect to perturbations,
in the sense that

\begin{Lemma}\label{cnbad}
If $(\lambda,\alpha)$ is  $(C,N)$ bad, then for any $C^{'}<C$, there exists $\epsilon=\epsilon(\lambda, \alpha,C-C^{'},N) >0$ such that if  $|\alpha^{'}-\alpha| < \epsilon$, then $(\lambda, \alpha^{'})$ is 
$(C^{'},N)$ bad.
\end{Lemma}

\begin{pf}
By H\"older continuity of the spectrum in Hausdorff topology \cite{AMS}, for any $E^{'} \in \Sigma(\lambda,\alpha^{'})$, there exists $E \in \Sigma(\lambda,\alpha)$,  with $|E-E^{'}|<C|\alpha^{'}-\alpha| ^{\frac{1}{2}}<C \epsilon ^{\frac{1}{2}}.$
Hence with estimate 
$$ | E- 2\lambda \cos2\pi (\theta+n\alpha)-E^{'}+ 2\lambda \cos2\pi(\theta+n\alpha^{'})| \leq C(  n\epsilon+ \epsilon ^{\frac{1}{2}} ),$$
 by a telescoping argument, we obtain that for any $\delta>0$, if $|m|>m_0(\lambda,\alpha,\delta)$ is large enough,   we have 
$$\sup_{\theta\in \R} \|A_m(E,\theta) -A^{'}_m(E^{'}, \theta)\| \leq e^{|m|(\ln \lambda +\delta)} (\epsilon^{\frac{1}{2}}+m \epsilon).$$
This is a  standard argument; see e.g. section 3 of  \cite{AYZ}  for details. 
Let $u_{E^{'}}^{\theta}(1)=u_E^{\theta}(1),
u_{E^{'}}^{\theta}(0)=u_E^{\theta}(0)$, where $u_E^{\theta}(k)$ is a
normalized solution of (\ref{ev}).  We have 
$$ | u_{E^{'}}^{\theta}(k)-u_E^{\theta}(k)|\leq (\epsilon^{\frac{1}{2}}+N  \epsilon)   e^{N(\ln \lambda +\delta)}  \qquad \forall |k|\leq N.$$
Therefore we can select $\epsilon$ small enough, not  depending on $\theta$, such that 
$$ \big(\sum_{|k|\leq N} |u_{E^{'}}^{\theta}(k)|^2 \big) ^{\frac{1}{2}}  \geq  \big(\sum_{|k|\leq N} |u_E^{\theta}(k)|^2 \big) ^{\frac{1}{2}} - \big(\sum_{|k|\leq N} |u_E^{\theta}(k)-u_{E^{'}}^{\theta}(k)|^2 \big) ^{\frac{1}{2}} > C^{'}.$$
The last inequality holds since $(\lambda,\alpha)$ is  $(C,N)$ bad. The smallness of $\epsilon$ is independent of  $E^{'} \in \Sigma(\lambda,\alpha^{'})$, since $N$ doesn't depend on $E \in \Sigma(\lambda,\alpha)$.
\end{pf}

For any given $\alpha\in \R\backslash \Q,$ $\varepsilon>0,$
$\beta^{'}>\ln \lambda> 0$, $K>0$, let
$\mathcal{S}(\alpha,\varepsilon,\beta^{'},K)$ be the set of $\alpha^{'}\in \R\backslash \Q$ that satisfy the following properties: $d_{H}(\alpha,\alpha^{'})<\varepsilon,$ $a_K(\alpha^{'})= e^{\beta^{'} q_{K-1}(\alpha^{'})}$, $\beta(\alpha^{'})=\beta^{'}$. The fundamental construction for our proof is the following:

\begin{Lemma}\label{construct}
Assume $\beta(\alpha)> \ln \lambda>0.$ Then for any $C>0$, $\beta^{'}>\ln \lambda> 0$, $\varepsilon>0,$ $M>0$, there exists $N=N(\lambda,\alpha)>0$, $K>M$ and $\alpha^{'} \in  \mathcal{S}(\alpha,\varepsilon,\beta^{'},K)$ such that $(\lambda,\alpha^{'})$ is  $(C,N)$ bad.
\end{Lemma} 

\begin{pf}
Let $\alpha= [a_1,a_2, \cdots]$. Since $\beta(\alpha)> \ln \lambda>0$, then by Proposition \ref{gordonayz},   for any given $C>0$, there exists $N\in\N$ such that $(\lambda,\alpha)$ is  $(C+1,N)$ bad.  We now construct $\alpha^{'}$ as follows. Let $\epsilon(\lambda, \alpha,C-C^{'},N) $ be as defined in Lemma \ref{cnbad}.  For any   $\beta^{'}>\ln \lambda> 0$, 
$\varepsilon>0,$ $M>0$, find  $K$ so that $K> \max\{
\frac{1}{\varepsilon},M\} $ and $q_{K-1}^{-2}<\epsilon(\lambda,
\alpha,1,N) .$ We define  $\alpha^{'}=[a_1^1,a_2^1,\cdots]$, where 
 \begin{eqnarray*}
a_n^1 = \left\{ \begin{array}{ccc}  a_n,  & n < K-1  \\
 e^{\beta^{'} q_{k^2 K-1}(\alpha)}, & n=k^2 K \\
1, &  k^2 K+1 \leq n \leq  (k+1)^2 K-1
\end{array} \right.
\end{eqnarray*}
By Lemma \ref{cnbad}, we have $(\lambda,\alpha^{'})$ is  $(C,N)$ bad,
and it is straightforward to check that 
$\alpha^{'} \in  \mathcal{S}(\alpha,\varepsilon,\beta^{'},K)$.
\end{pf}

We now finish the proof of Theorem \ref{main}(1). For any  $\alpha\in \R\backslash \Q$ with $\lambda=e^{\beta(\alpha)}$, we write 
$\alpha= [a_1,a_2, \cdots].$ For any $\varepsilon>0$, by the
construction of Lemma \ref{construct}, one can find $\alpha^{1}$ such that $\alpha^{1} \in  \mathcal{S}(\alpha, \frac{\varepsilon}{2}, \ln \lambda+\frac{1}{2},  N^1)$ with $N^1=[ \frac{2}{\varepsilon}].$  This implies that 
  $\beta(\alpha_{1})= \ln \lambda+\frac{1}{2}$. Then by Proposition \ref{gordonayz},   for given $C_1>0$, there exists $N_1\in\N$ such that $(\lambda,\alpha^{1})$ is  $(C_1,N_1)$ bad.  
 We now construct $\alpha^{k}$ by induction.  Given $\alpha^{k}$, $C_k=k C_1$,  
 by Lemma \ref{construct}, there exist $\alpha^{k+1} \in  \mathcal{S}(\alpha^k, \frac{\varepsilon}{2^{k+1}},  \ln \lambda+ \frac{1}{2^{k+1}}, N^{k+1})$, $N_{k
 +1}=N(\lambda,\alpha_k)$, $N^{k+1}> 2 N^{k}$ such that $(\lambda,\alpha^{k+1})$ is  $(C_{k+1}, N_{k+1})$ bad. Let
 $\alpha_{\infty}=\lim \alpha^{k}$. The limit exists since $d_H(\alpha^{k},\alpha^{k+1})< \frac{\varepsilon}{2^k}$. Also by the construction, we have 
 \begin{eqnarray}\label{cons1}a_{N^k}(\alpha_{\infty})&=& e^{(\ln\lambda+ \frac{1}{2^{k}}) q_{N^k-1}(\alpha_{\infty})},\\
 \label{cons2} |\alpha_{\infty} -\alpha^k | &<&  ( \frac{1}{q_{N^k-1}(\alpha_{\infty})})^2<\epsilon(\lambda, \alpha^{k},1,N_k).\end{eqnarray} 
 Therefore $(\ref{cons1})$ implies that $\beta(\alpha_{\infty})=\ln \lambda$. By Lemma \ref{cnbad} and $(\ref{cons2})$, we have $(\lambda, \alpha_\infty)$ is $(C_k-1,N_k)$ bad, so $H_{\lambda,\alpha_\infty,\theta}$ has
purely singular continuous spectrum for all $\theta$ by Lemma \ref{rem}.\qed

\section{Pure point spectrum}

The pure point spectrum will be a corollary of the following full measure reducibility result. 

\begin{Theorem}\label{inftyredu}
Let $0<\beta(\alpha)<\infty.$ There exists a dense set of $\alpha$
with $\lambda=e^{\beta(\alpha)}$, such that for a full measure set of $E \in \Sigma(H_{\lambda^{-1},\alpha})$, the almost Mathieu cocycle $(\alpha, S_E^{\lambda^{-1}})$ is $C^\infty$ reducible.
\end{Theorem}

\textbf{Proof of Theorem \ref{main}(2)} 
From \cite{AYZ}, one knows that if $(\alpha, S_E^{\lambda^{-1}})$ is
$C^\infty$ reducible for a full measure set of $E \in
\Sigma(H_{\lambda^{-1},\alpha})$, then for almost every $\phi$,
$H_{\lambda,\alpha,\phi}$ has pure point spectrum. See also Theorem 3.1 of \cite{JK} for a more general result of this nature. Thus Theorem \ref{main}(2) follows from Theorem \ref{inftyredu}.\qed
\\

To prove  Theorem \ref{inftyredu}, we start with

\begin{Lemma}\label{keyredu}
Let $ \lambda>1$, $0<\beta<\infty,$ $0<\delta_j<\beta/2$, $\tau>1,\gamma_1,\gamma_2>0$.  Suppose that  $\alpha_j= [ a_1, \cdots, a_j, 1,1, \cdots ]$ and let 
$\alpha_{j,n}=[\tilde{a}_1,\tilde{a}_2,\cdots]$, where 
 \begin{eqnarray*}
\tilde{a}_i = \left\{ \begin{array}{ccc}   a_i ,  & i \leq n-1  \\
 e^{(\beta -2\delta_j)q_{n-1}(\alpha_j) }, & i=n \\
1, &  i \geq n+1
\end{array} \right.
\end{eqnarray*}
 Suppose that $\rho(\alpha_j, S_{E}^{\lambda^{-1}})\in
 DC_{\alpha_j}(\gamma_1,\tau)$,    $\rho(\alpha_{j,n},
 S_{E}^{\lambda^{-1}}) \in DC_{\alpha_{j,n}}(\gamma_2,\tau)$.   Then
 there exist $B_j \in C^\omega( \T,
 SL(2,\R))$ and $B_{j,n}\in C^*( \T,
 SL(2,\R)),$ with $*=\omega$ if $\ln\lambda>\beta$ and
 $\infty$ if $\ln\lambda=\beta,$ so that
 $(\alpha_j, S_{E}^{\lambda^{-1}})$ is reducible by $B_j(\theta)$ and
 $(\alpha_{j,n}, S_{E}^{\lambda^{-1}})$ is reducible by
 $B_{j,n}(\theta).$ Moreover,
 for any $\epsilon>0$, 
  there exists  $N=N(\alpha_j, \delta_j, \tau, \gamma_1,\gamma_2,\epsilon)$, such that for $n\geq N$, we have the following:
\begin{enumerate}
\item if $\lambda>e^{\beta}$, then  $\|B_{j,n}-B_j\|_{\ln\lambda -\beta} \leq \epsilon$.
\item if $\lambda=e^{\beta}$, then  $dist_{C^\infty}(B_{j,n},B_j) \leq \epsilon$.
\end{enumerate}
\end{Lemma}

\begin{pf}

We first recall

\begin{Theorem}\label{al}\cite{J99}
Suppose that $\alpha$ is Diophantine, $\theta$ is Diophantine w.r.t
$\alpha$, $\lambda>1.$ Then $H_{\lambda,\alpha, \theta}$ has pure point spectrum with exponentially decaying eigenfunctions. Moreover, each eigenfunction $u(n)$ satisfies 
\begin{eqnarray}\label{eigen} \lim_{|n|\rightarrow \infty}  \frac{\ln (u^2(n)+u^2(n+1) )}{2|n|}=-\ln \lambda.
\end{eqnarray}
\end{Theorem}

By the definition of  $\alpha_j$, we know it is Diophantine. Then by
the assumption that  $\lambda>1$, $\rho(\alpha_j,
S_{E}^{\lambda^{-1}})\in DC_{\alpha_j}(\tau,\gamma_1)$,  using Theorem
\ref{al} and Aubry duality (e.g. \cite{AJ08}), we have the following result: 
for any given  $\delta_j>0$,  there exist $T_j=T_j(\alpha_j,\delta_j,\gamma_1,\tau)$,  $B_j(\theta)\in C^\omega_{\ln\lambda- \delta_j/4}( \T, SL(2,\R))$, such that  
\begin{eqnarray}\label{red}B_j(\theta+\alpha_j) S_{E}^{\lambda^{-1} }(\theta) B_j(\theta)^{-1}= R_{\rho(\alpha_j, S_E^{\lambda^{-1}})},\end{eqnarray}
$deg B_j(\theta)=0$, and   \begin{eqnarray}\label{C}\|B_j\|_{\ln
  \lambda-\delta_j/4} \leq T_j.\end{eqnarray}
Theorem 2.5 of \cite{AJ08} contains
the proof  of  this standard result; we just point out
that the fact that $B_j(\theta)\in C^\omega_{\ln\lambda- \delta_j/4}( \T, SL(2,\R))$ follows from $(\ref{eigen})$.

By $(\ref{red})$ and the Cauchy estimate, we have
$$B_j(\theta+\alpha_{j,n}) S_{E}^{\lambda^{-1}} (\theta) B_j(\theta)^{-1}=e^{F_j(\theta)} R_{\rho(\alpha_j,S_E^{\lambda^{-1}})},$$ 
with estimate \begin{eqnarray*}\|F_j\|_{ \ln\lambda- \delta_j/2} \leq |\alpha_j-\alpha_{j,n}|\|\partial B_j\|_{\ln\lambda- \delta_j/2} \|B_j\|_{\ln\lambda-  \delta_j/4}\leq \frac{4T_j^2}{q_{n-1}^2 \delta_j}.\end{eqnarray*}
Then there exists $\tilde{N}=\tilde{N}(\alpha_j, \delta_j, \tau, \gamma_1,\gamma_2)$, such that if $n\geq \tilde{N}$, 
then we have 
\begin{eqnarray}\label{small}
\|F_j\|_{ \ln\lambda- \delta_j/2}  \leq \frac{4T_j^2}{q_{n-1}^2 \delta_j} \leq \varepsilon(\tau,\gamma_2, \delta_j/2),
\end{eqnarray}
 where $\varepsilon=\varepsilon(\tau,\gamma,  h-h' )$ is defined in Theorem \ref{hy1}. 

Since $\rho(\alpha_{j,n}, S_{E}^{\lambda^{-1}})\in DC_{\alpha_{j,n}}(\tau,\gamma_2)$, $\deg B_j(\theta)=0$, we have 
$$\rho(\alpha_{j,n},e^{F_j(\theta)} R_{\rho(\alpha_j,S_E^{\lambda^{-1}})})=\rho(\alpha_{j,n}, S_{E}^{\lambda^{-1}}) \in DC_{\alpha_{j,n}}(\tau, \gamma_2),$$
so by $(\ref{small})$, one can apply
Theorem \ref{hy1}, getting  $\widetilde{B}_n(\theta)\in C^\omega_{\ln\lambda- \delta_j}( \T, SL(2,\R))$, $\varphi_n \in C^\omega_{\ln\lambda- \delta_j}( \T,\R)$, such that 
$$\widetilde{B}_n(\theta+\alpha_{j,n})e^{F_j(\theta)} R_{\rho(\alpha_j,S_E^{\lambda^{-1}})}  \widetilde{B}_n(\theta)^{-1}= R_{\varphi_n(\theta)}.$$
Moreover, we have  the estimates 
\begin{eqnarray} \label{A}
\|\widetilde{B}_n-\id\|_{\ln\lambda- \delta_j} &\leq&\frac{cT_j}{q_{n-1} \delta_j^{1/2}},\nonumber\\
\|\varphi_n(\theta)-\hat\varphi_n(0)\|_{\ln\lambda- \delta_j} &\leq& \frac{cT_j^2}{q_{n-1}^2 \delta_j}.
\end{eqnarray}

Now we
let  $$\psi_n(\theta)-\psi_n(\theta+\alpha_{j,n})=\varphi_n(\theta)-\hat{\varphi}_n(0),$$ so we have \begin{eqnarray} \label{B}  \hat{\psi}_n(k)=  \frac{  \hat{\varphi}_n(k) }{1-e^{2\pi i \alpha_{j,n}}}.\end{eqnarray}
Set $\widetilde{q}_n=q_n(\alpha_{j,n})$. Then by
(\ref{A}),(\ref{B}),(\ref{appro2}), and Lemma \ref{bdc},  we have  the following estimates:
\begin{eqnarray*}
\|\psi_n\|_{\ln \lambda -\beta} &\leq& \Big(   \sum_{0< |k|<q_{n-1}}+ \sum_{ q_{n-1}\leq |k|< \widetilde{q}_n }+  \sum_{ |k|\geq \widetilde{q}_n}  \Big) \frac{|\hat{\varphi}_n(k)|}{\|k\alpha_{j,n}\|_{\T}}e^{|k|(\ln\lambda-\beta)}\\
&\leq &  c\Big(q_{n-1}+ q_{n-1} e^{q_{n-1} (\beta-  2\delta_j) }  \sum_{ q_{n-1}\leq |k|< \widetilde{q}_n } e^{-|k| (\beta-  \delta_j )} \\
&&+ \sum_{ |k|\geq \widetilde{q}_n}  4|k|  e^{-|k| (\beta  - \delta_j )}\Big) \frac{T_j^2}{q_{n-1}^2 \delta_j} \leq \frac{cT_j^2}{q_{n-1} \delta_j}
\end{eqnarray*}
which implies that $\psi_n\in C^\omega_{\ln\lambda-\beta}(
\T,\R)$. Let
$B_{j,n}(\theta)=R_{\psi_n(\theta)}\widetilde{B}_n(\theta)
B_j(\theta)$. Then by the definition of $\psi_n(\theta)$, we have 
$$B_{j,n}(\theta+\alpha_{j,n}) S_{E}^{\lambda^{-1} }(\theta) B_{j,n}(\theta)^{-1}= R_{\hat{\varphi}_n(0)}.$$

Moreover,  for any $\epsilon>0$, there exists $N=N(\alpha_j, \delta_j, \tau, \gamma_1,\gamma_2,\epsilon)$, such that if $n\geq N$, we have \begin{eqnarray*}
\|B_j-B_{j,n}\|_{\ln\lambda -\beta} &\leq& 2\|B_j\|_{\ln \lambda-\delta_j/4} ( \|\tilde{B}_n-id\|_{\ln \lambda-\delta_j} + \|R_{\psi_n}-id\|_{\ln\lambda -\beta})\\
&\leq& \frac{cT_j^3}{q_{n-1} \delta_j} \leq \epsilon,
\end{eqnarray*}
which establishes the first part of the lemma.

 For the second part, note that for $s\geq 0,$ we have
 \begin{eqnarray*}
 \|f\|_{C^s} \leq \sum_{j=0}^s \sum_{k\in \Z}  |k|^j | \hat{f}(k)|. 
 \end{eqnarray*}
 Then, similar as before,  for any $s\in\Z$, we have 
 \begin{eqnarray*}
&&\|\psi_n\|_{C^s} \\ &\leq& \Big(   \sum_{0< |k|<q_{n-1}}+ \sum_{ q_{n-1}\leq |k|< \widetilde{q}_n }+  \sum_{ |k|\geq \widetilde{q}_n}  \Big) \frac{|k|^{s+1} |\hat{\varphi}_n(k)|}{\|k\alpha_{j,n}\|_{\T}} \\
&\leq&  \Big(q_{n-1}\sum_{0< |k|<q_{n-1}}   |k|^{s+1} e^{-|k| (\beta-  \delta_j )} \\
&& + q_{n-1}e^{q_{n-1} (\beta-  2\delta_j) }  \sum_{ q_{n-1}\leq |k|< \widetilde{q}_n } 
 |k|^{s+2} e^{-|k| (\beta-  \delta_j )} \\
&&+ \sum_{ |k|\geq \widetilde{q}_n}  4|k|^{s+2}  e^{-|k| (\beta  - \delta_j )}\Big) \frac{T_j^2}{q_{n-1}^2 \delta_j} \leq \frac{c_sT_j^2}{q_{n-1} \delta_j}.
\end{eqnarray*}
Therefore one can find $N=N(\alpha_j, \delta_j, \tau, \gamma_1,\gamma_2,\epsilon,s)$, such that if $n\geq N$, one reaches 
$\|B_j-B_{j,n}\|_{C^s} \leq 2\|B_j\|_{\ln \lambda-\delta_j/4} ( \|\tilde{B}_n-id\|_{\ln \lambda-\delta_j/2} + \|R_{\psi_n}-id\|_{C^s}) \leq \epsilon/2.
$
Choosing $N=\max_{s\leq C\ln \epsilon^{-1}} N(\alpha_j, \delta_j, \tau,
\gamma_1,\gamma_2,\epsilon,s),$ we obtain the desired result.

\end{pf}

\begin{Lemma}\label{derrot}
Let $\lambda>1$, $\alpha\in \R\backslash \Q.$ For a full
measure set of $E \in \Sigma(H_{\lambda^{-1},\alpha})$, there exists $B_E \in C^\omega(\T, SL(2,\R))$ such that
$$   B_E(\theta+\alpha) S_E^{\lambda^{-1}}(\theta)B_E(\theta)^{-1}\in SO(2,\R).$$
Furthermore,  we have 
\begin{equation}\label{der}
\frac{d \rho(\alpha, S_{E}^{\lambda^{-1}})}{d E}= -\frac{1}{8\pi} \int_{\T} \|B_E(\theta)\|_{HS}^2 d\theta \leq -\frac{1}{4\pi}.
\end{equation}
\end{Lemma}
Here $\|\cdot\|_{HS}$ denotes the Hilbert-Schmidt norm.
\begin{pf}
This is a combination of full measure rotations reducibility \cite{AFK,YZ} and formula $(1.5)$ of \cite{AFK}.   
\end{pf}

\textbf{Proof of Theorem \ref{inftyredu}}
Let $\alpha\in \R\backslash \Q$ with $\lambda=e^{\beta(\alpha)}$. We write 
$\alpha= [a_1, a_2,\cdots].$ 
                             Assume $\beta=\beta(\alpha)=\ln\lambda$.  
For any $\varepsilon>0$, we  first perturb $\alpha$ to $\alpha_0=[
a_1, \cdots, a_{n_0}, 1, 1, \cdots ]$ so that $d_H(\alpha, \alpha_0)=
\frac{1}{n_0+1}< \frac{\varepsilon}{2}$. Fix $4\delta_0< \beta,$
  $\tau>1,\gamma>0$. We now proceed  by induction. Given
  $\alpha_{j-1}$, $j\geq 1,$ find  $$n_j=\max\{N( \alpha_{j-1}, \frac{\delta_0}{2^{j-1}},\tau, \frac{\gamma}{2^{j-1}},\frac{\gamma}{2^j}, \frac{\varepsilon}{2^j}), n_{j-1}+1,  \frac{ 2^j}{\epsilon}\},$$ 
  where $N=N(\alpha_j, \delta_j, \tau,\gamma_1,\gamma_2,\epsilon)$ is as defined in Lemma \ref{keyredu}.  Then define  $\alpha_j=[a_1^j,a_2^j,\cdots],$ where
  \begin{eqnarray*}
a_i^{j} = \left\{ \begin{array}{ccc}   a_i^{j-1},  & i<n_j  \\
e^{(\beta-\frac{2\delta_0}{2^{j-1}})q_{n_j-1}(\alpha_{j-1})}, & i=n_j \\
1, &  i \geq n_j+1
\end{array} \right.
\end{eqnarray*}
 Set $\alpha_{\infty}=\lim \alpha_j$. The limit exists since $d_H(\alpha^{j},\alpha^{j-1})< \frac{\varepsilon}{2^j}$.  Also by the construction, for $j\geq 1$, we have $$a_{n_j} (\alpha_{\infty}) =e^{(\beta-\frac{2\delta_0}{2^{j-1}})q_{n_j-1}(\alpha_{\infty})},$$ which implies that 
 $\beta(\alpha_{\infty})=\beta$. For these $\alpha_j$, we define
$$\mathcal{B}= \bigcup_{n=1}^{\infty}\bigcap_{j=n} ^{\infty}\{E| \rho(\alpha_j, S_{E}^{\lambda^{-1}}) \in DC_{\alpha_j}(\frac{\gamma}{2^{j}},\tau)\}.$$
By Lemma \ref{derrot}, we have  $Leb(\{E| \rho(\alpha_j,
S_{E}^{\lambda^{-1}}) \notin
DC_{\alpha_j}(\frac{\gamma}{2^{j}},\tau)\})=O( \frac{\gamma}{2^{j}})$
uniformly in $j$, thus $\mathcal{B}$ is a full measure set by the
Borel-Cantelli Lemma. For any fixed $E\in \mathcal B$, there exists
$n_E\in\Z$ such that $ \rho(\alpha_j, S_{E}^{\lambda^{-1}})\in
DC_{\alpha_j}(\frac{\gamma}{2^{j}},\tau)$ for any $j\geq n_E$. By the
definition of $\alpha_j$ and Lemma \ref{keyredu}, we know
$(\alpha_j,S_E^{\lambda^{-1}})$ is reducible by $B_j\in C^\infty(\T,
SL(2,\R))$, and $dist_{C^\infty}(B_j,B_{j-1})<
\frac{\epsilon}{2^j}$. Let $B_{\infty} (\theta)= \lim B_j(\theta)$,
then $B_{\infty} \in C^\infty(\T, SL(2,\R))$ and
$(\alpha_{\infty},S_E^{\lambda^{-1}})$ is reducible by $B_{\infty},$
so $C^{\infty}$ reducible. \qed

\section*{Acknowledgements} A.A. and Q.Z. were partially
supported by the ERC Starting Grant\\\textquotedblleft
Quasiperiodic\textquotedblright. S.J. was a 2014-15 Simons
Fellow, and  was partially supported by NSF DMS-1401204. Q.Z. would
also like to thank the hospitality of the UCI where this work was
started. S. J. and Q. Z. are grateful to the Isaac
    Newton Institute for Mathematical Sciences, Cambridge, for its hospitality supported by EPSRC Grant Number EP/K032208/1, during the programme \textquotedblleft
Periodic and Ergodic Spectral
    Problems\textquotedblright  where they worked on this paper.

\end{document}